\documentclass[11pt,twoside]{article} 
\usepackage{asp2004}
\usepackage{epsf}
\usepackage{psfig}
\usepackage{lscape} 

\begin{document} 
\title{Photometric Variability in the Faint Sky Variability Survey}
\author{L. Morales-Rueda,$^1$ P. J. Groot,$^1$ T. Augusteijn,$^2$
G. Nelemans,$^3$ P. M. Vreeswijk,$^4$ and E.J.M. van den
Besselaar$^1$} \affil{$^1$Department of Astrophysics, University of
Nijmegen, P.O. Box 9010, 6500 GL Nijmegen, The Netherlands\\$^2$Nordic
Optical Telescope, Ap. 474, 38700, La Palma, Spain\\ $^3$Institute of
Astronomy, Madingley Rd, CB3 0HA Cambridge, UK\\$^4$Astronomical
Institute Anton Pannekoek, 1098 SJ Amsterdam, The Netherlands.}

\begin{abstract} 
The Faint Sky Variability Survey (FSVS) is aimed at finding
photometric and/or astrometric variable objects between 16th and 24th
mag on time-scales between tens of minutes and years with photometric
precisions ranging from 3 millimag to 0.2 mag. An area of $\sim$23
deg$^2$, located at mid and high Galactic latitudes, was covered using
the Wide Field Camera (WFC) on the 2.5-m Isaac Newton Telescope (INT)
on La Palma. Here we present some preliminary results on the
variability of sources in the FSVS.
\end{abstract}

\section{Introduction}
The FSVS data set consists of 78 INT WFC fields. Each field is divided
into four, corresponding to the 4 2kX4k CCDs that form the WFC. For
each field, we took one set of B, I and V images on a given
night. Several more images were taken in the V band on that night and
on consecutive nights. Typically, fields were observed in V 10 -- 20
times within one week. Exposure times were 10 min with a dead time
between observations of 2 min.  This observing pattern allows us to
sample periodicity timescales from 2$\times$(observing time + dead
time) (i.e. 24 min) up to the maximum time span of the observations
(i.e. a few days). The combination of the colour information and the
variability allows us to distinguish between different types of
variable systems. See Groot et al. (2003) for a full discussion on the
FSVS data.

\section{The Floating Mean Periodogram}
In order to combine the colour with the variability information
contained in the FSVS we must be able to measure the variability
timescale associated with each target. Because of the relatively few
number of V observations per field (between 10 and 20) we use the
``floating mean'' periodogram technique to obtain the characteristic
variability timescale in each case. This method has been used
successfully in planet searches (Cumming et al. 1999) and to determine
the orbits of subdwarf B binaries (Morales-Rueda et al. 2003).

The floating mean periodogram consists of fitting the data with a
model composed of a sinusoid plus a constant of the form
$\gamma + K \sin(2 \pi f (t - t_0)),$
where $f$ is the frequency and $t$ is the observation time. This
method corrects a failing of the Lomb-Scargle periodogram (Lomb 1976;
Scargle 1982) which subtracts the mean of the data and then fits a
sinusoidal, which is incorrect for small numbers of points.

To test whether the periodogram was able to recover the correct
periods, as a first step, we simulated the brightness variations of a
series of sources with variability periods between 24 min and the time
span of the observations, using the time sampling of {\em one of the
fields}, and generated fake lightcurves for each period. We used the
floating mean periodogram to calculate the most probable variability
timescale of each lightcurve, and plotted the calculated versus the
real period. The result is a linear curve with a 45$^{\circ}$ slope
(i.e. real periods are recovered successfully) and a complicated error
structure. Examples of this test for two given variability amplitudes
are presented in Fig. 1. Using a different time sampling (which is
equivalent to analysing a different field) will generate a slightly
different graph.

\begin{figure}[!ht]
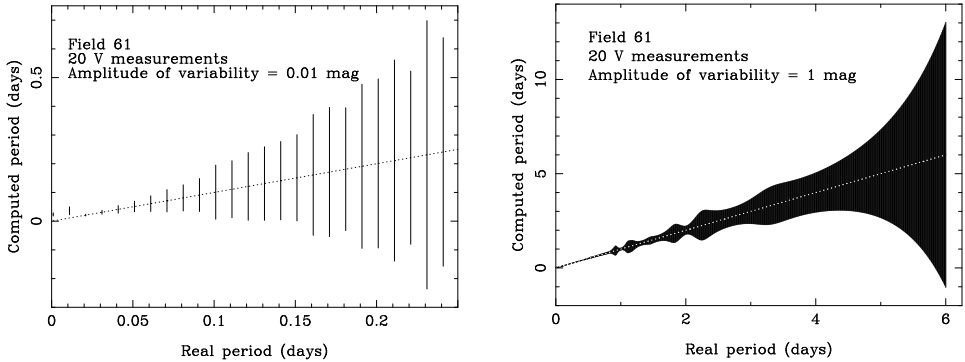

\plotfiddle{compf61amp001mp025.ps}{3.2cm}{-90}{27}{27}{-200}{125}
\plotfiddle{compf61amp1mp6.ps}{0cm}{-90}{27}{27}{-10}{150}
\caption{The variability timescale calculated using the floating mean
  periodogram is plotted versus the real variability timescale of data
  simulated using the time sampling of one of the FSVS fields. Data
  for two different variability amplitudes is presented.}
\end{figure}

A more systematic and complete test to check that the method works,
and to calculate its period detection efficiency, consists in
generating curves like those of Fig. 1 for a complete range of values
of variability timescale and amplitude, and object brightness (which
is directly related to the uncertainty in the brightness measured for
each band). These calculations must be done for each time sampling in
the data (Morales-Rueda et al. in preparation).

\section{Results}
When we run the floating mean periodogram on the real data lightcurves
we obtain their most likely variability timescale and the amplitude of
that timescale. In the following colour-colour diagrams we show the
point sources from the FSVS that show no variability (top left panel),
the point sources that show variability (top right panel) and the
ratio between variable and non-variable point sources (bottom
panel). Variability is determined in each case by calculating the
$\chi^2$ of the light curve with respect to its average value. Objects
with $\chi^2$ above the 5-$\sigma$ variability level are considered
variable (Groot et al. 2003). The ratio is presented in percentages.
The main difference between the distribution of variable and
non-variable objects is the excess of variable systems at colours
B$-$V$\sim$0 and V$-$I$\sim$1. These sources will be mostly QSOs but
we also expect any cataclysmic variables (CVs) in the field to appear
in that colour-colour region. The fraction of variable to non-variable
objects along the main sequence seems fairly uniform, but more
rigorous calculations are required before we can draw any conclusions.

\begin{figure}[!ht]
\plotfiddle{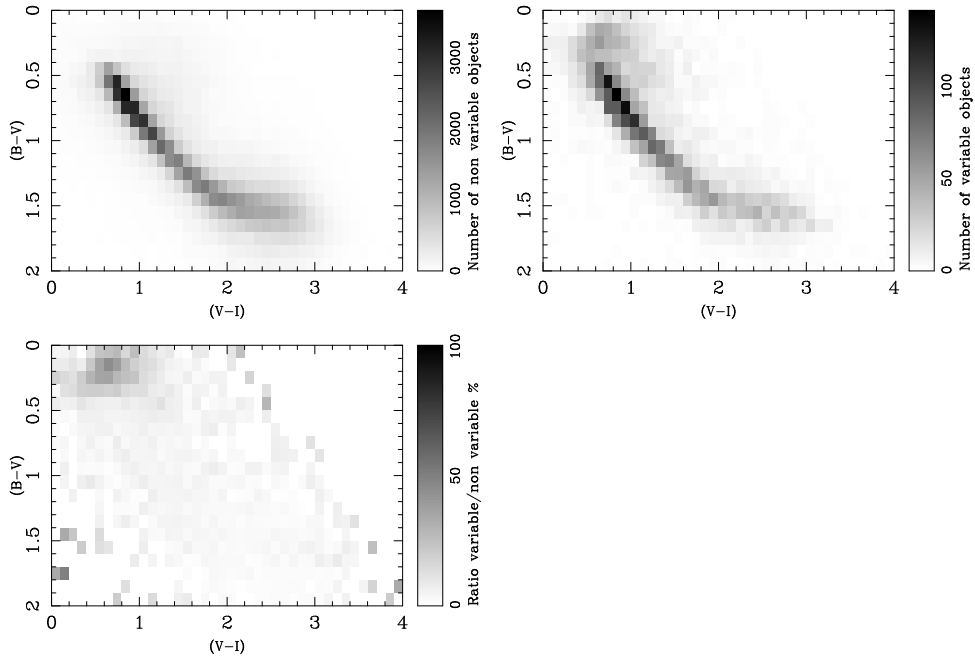}{7.8cm}{-90}{50}{50}{-200}{270}
\caption{Colour-colour diagrams showing the number of non-variable
  (top left panel) and variable (top right panel) systems found in the
  FSVS and the fraction of variable to non-variable systems (bottom
  panel).}
\end{figure}

We also display the results in colour-colour diagrams for different
ranges of variability timescales and amplitudes. The ranges have been
selected so the first would include orbital periods corresponding to
CVs (variability scales up to 6 hours), the second RR Lyr stars
(scales from 6 hours to 1 day), the third longer variability trends of
CVs (from 1 to 4 days), and anything else (above 4 days). Four
different amplitude ranges are also chosen. In all colour-colour
diagrams we have also plotted the 3-$\sigma$ upper limit of the main
sequence for clarity. We find no obvious correlation between the time
scales of the variability and their amplitudes.

\begin{figure}[!ht]
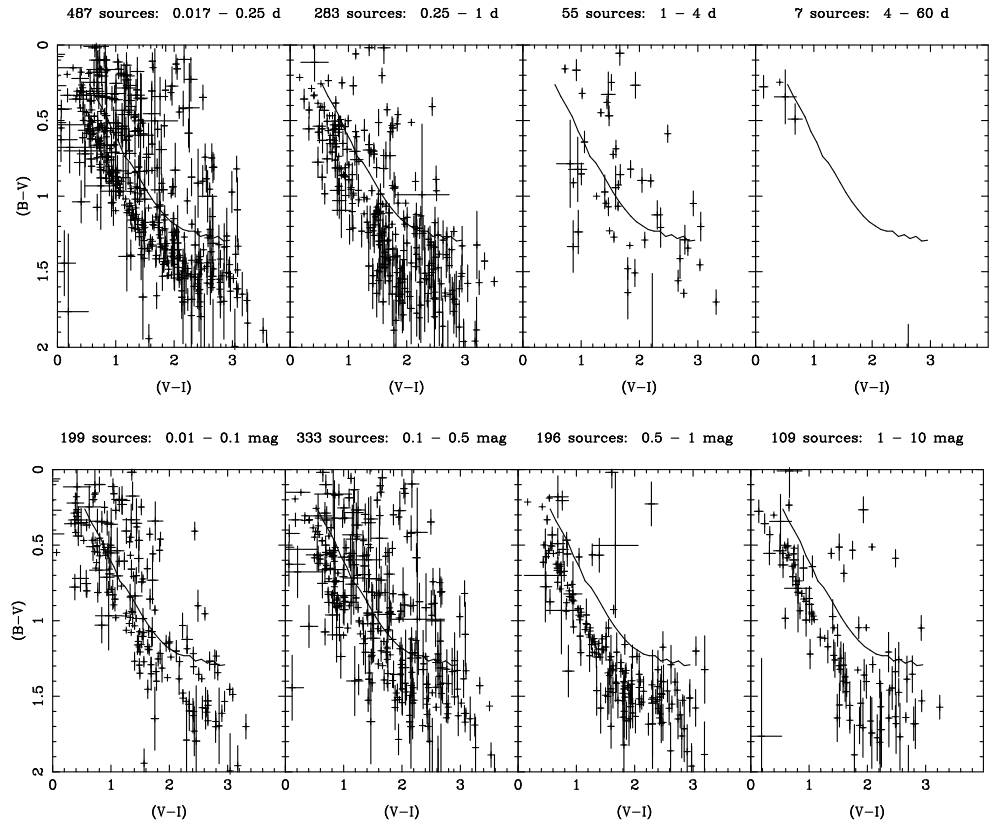


\plotfiddle{cmdp.ps}{4.8cm}{-90}{58}{58}{-230}{165}
\plotfiddle{cmda.ps}{4.3cm}{-90}{58}{58}{-230}{140}
\vspace{1cm}
\caption{Colour-colour diagrams for four different variability (top
  panel) and amplitude (bottom panel) ranges. Note that some sources
  lie outside the ranges plotted.}
\end{figure}

In a preliminary analysis of the FSVS we find that, down to 24 mag,
there are of the order of 500 objects in the variability range
corresponding to CVs and 300 in the range corresponding to RR Lyr
stars. We find 62 sources showing longer variability periods. The
number of point sources found in each variability and amplitude range
is given in Fig.~3. These preliminary values have not been corrected
for the period detection efficiency discussed in Section 2 which might
alter the results considerably.

\acknowledgements{We thank T. R. Marsh for making his software
  available. The FSVS was supported by NWO Spinoza grant 08-0. The
  FSVS is part of the INT Wide Field Survey.  LMR, EJMvdB and PJG are
  supported by NWO-VIDI grant 63g.ou2.201.  The INT is operated on the
  island of La Palma by the Isaac Newton Group in the Spanish
  Observatorio del Roque de los Muchachos of the Instituto de
  Astrof\'isica de Canarias.}

\end{document}